\documentclass[10pt,twocolumn]{article}
\pdfoutput=1
\usepackage{authblk}
\usepackage[english]{babel}
\usepackage{graphicx}
\usepackage{url}
\usepackage{subfig}

\usepackage{xcolor}

\usepackage{amsfonts}
\usepackage{amssymb}
\usepackage{amsmath}

\title{The Zen of Multidisciplinary Team Recommendation}
\author{Anwitaman Datta}
\author{Stefano Braghin}
\author{Jackson Tan Teck Yong}
\affil{School of Computer Engineering\\Nanyang Technological
University,
Singapore\\\texttt{\{anwitaman,s.braghin,jacktty\}@ntu.edu.sg}}

\date{}

\begin{document}
\maketitle

\begin{abstract}
In order to accomplish complex tasks, it is often necessary to compose
a team consisting of experts with diverse competencies. However, for
proper functioning, it is also preferable that a team be socially
cohesive. A team recommendation system, which facilitates the search
for potential team members can be of great help both for (i) individuals who need
to seek out collaborators and (ii) managers who need to build a team
for some specific tasks.

A decision support system which readily helps summarize such metrics,
and possibly rank the teams in a personalized manner according to the
end users' preferences, can be a great tool to navigate what would
otherwise be an information avalanche. 

In this work we present a general framework of how to compose such
subsystems together to build a composite team recommendation system, and instantiate it for a case study of academic teams.

\end{abstract}

\section{Introduction}
\label{sec:introduction}

In order to accomplish complex tasks, it is often necessary to compose
a team consisting of experts with diverse competencies. However, for
proper functioning, it is also preferable that a team be socially
cohesive. A team recommendation system, which facilitates the search
for potential team members, as well as allow profiling specific team
configuration can be of great help both for (i) individuals who need
to seek out collaborators and (ii) managers who need to build a team
for some specific tasks.

While there is arguably no well-defined notion of a ``best team'', one
may quantify the quality of a team according to multiple metrics - and
a decision support system which readily helps summarize such metrics,
and possibly rank the teams in a personalized manner according to the
end users' preferences, can be a great tool to navigate what would
otherwise be an information avalanche.

A team recommendation system needs to build upon many smaller
subsystems, many of which are subjects of study on their own right -
for instance, expertise identification for individuals, topic
extraction from documents, multi-dimensional social network modeling
and analysis, graph mining and identification of implicit relations, etc.

In this work we present a general framework of how to compose such
subsystems together to build a composite team recommendation system.
In fact, the trait that mostly differentiate our framework from
several approaches to team recommendation is that SWAT (Social Web
Application for Team Recommendation) is a general approach while some
existing solutions focus on specific sub-problems, like expert finding
(SmallBlue \cite{smallblue}), implicit relation identification
(WikiNetViz \cite{DBLP:conf/isi/LeDLD08}), etc.

Following the general framework, we will then discuss a specific
case study (instantiation of the framework) of team-recommendation
for scientific collaboration (this is relatively easy to realize
given the abundance of publicly available information), and describe
a system which is implemented as a stand-alone as well as Facebook
integrated application to harness information from multiple sources
including prominent bibliographic databases and the Facebook network
itself. We also discuss a few collaboration enabling features that
have in addition been integrated in the system, on top of the
primary task of recommendation.


\section{Model}
\label{sec:model}
\begin{figure}[htbp]
   \centering
   \includegraphics[width=.95\columnwidth]{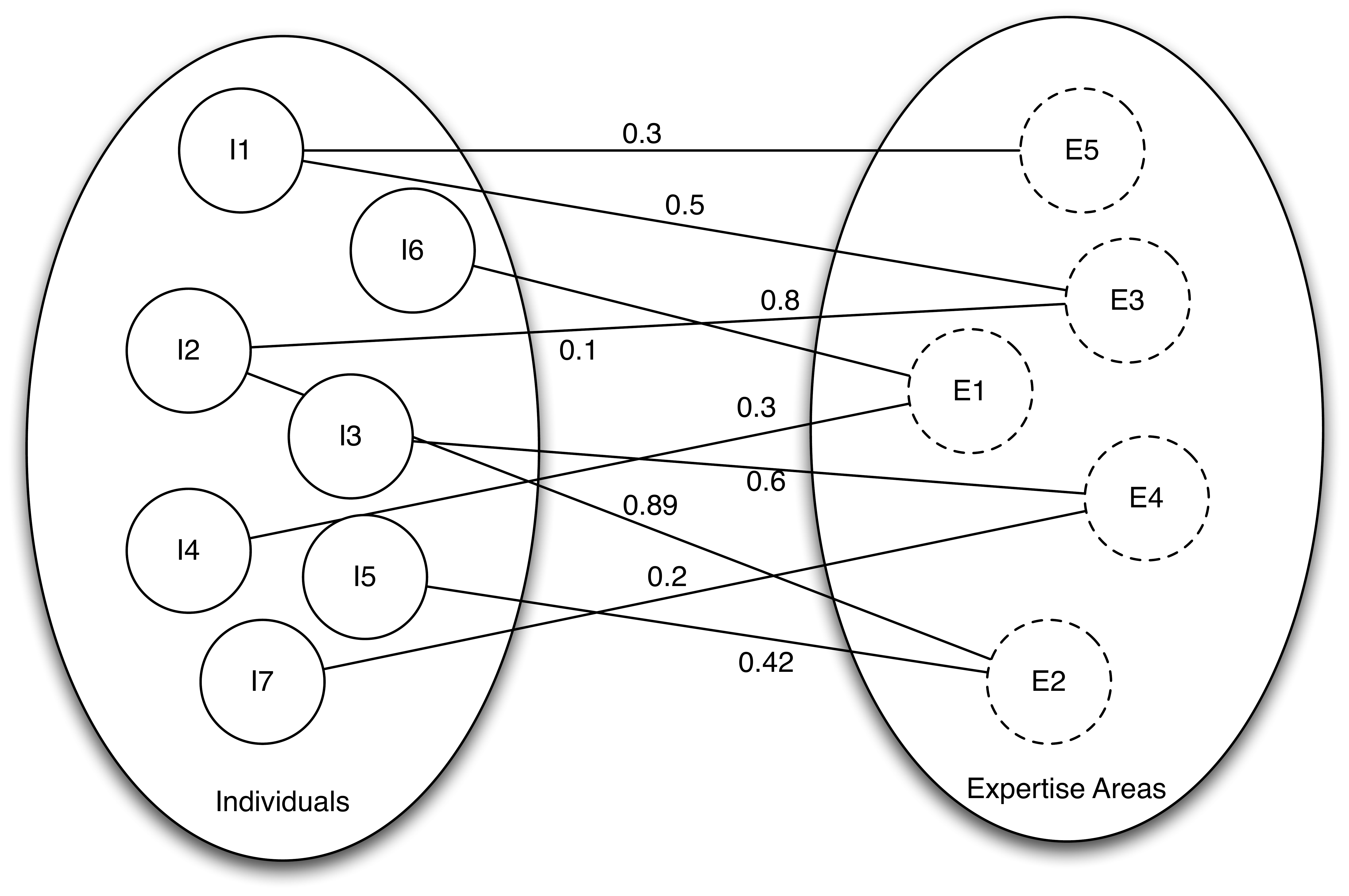}
 \caption{The \emph{competence graph} connects each individual to
     the expertise areas in which she/he has a certain level of
     competence. The competence level is expressed by means of a label
     associated with each edge of the graph.}
 \label{fig:competence}
\end{figure}

\begin{figure}[htbp]
  \centering
  \includegraphics[width=.85\columnwidth]{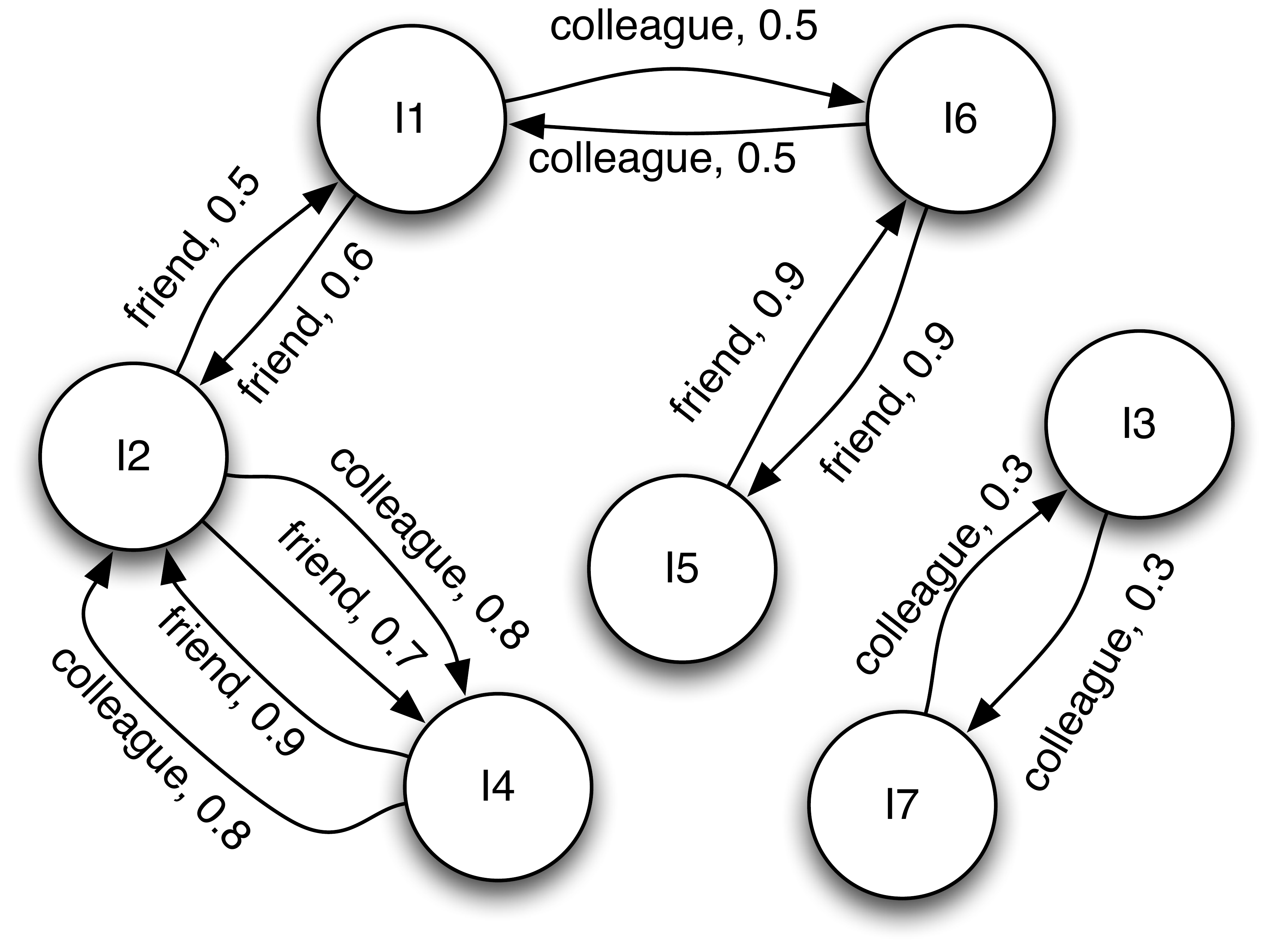}
\caption{The \emph{social graph} represents the relationships
    existing among the individuals of the system. Such relationship is categorized by a type -- which identifies the social dimension expressed by the edge -- and a numeric value -- which represents the strength of the relationship. The semantics of the relationship type may be application dependent, and utilized accordingly.}
\label{fig:social}
\end{figure}

\begin{figure}[htbp]
\centering
\includegraphics[width=.95\columnwidth]{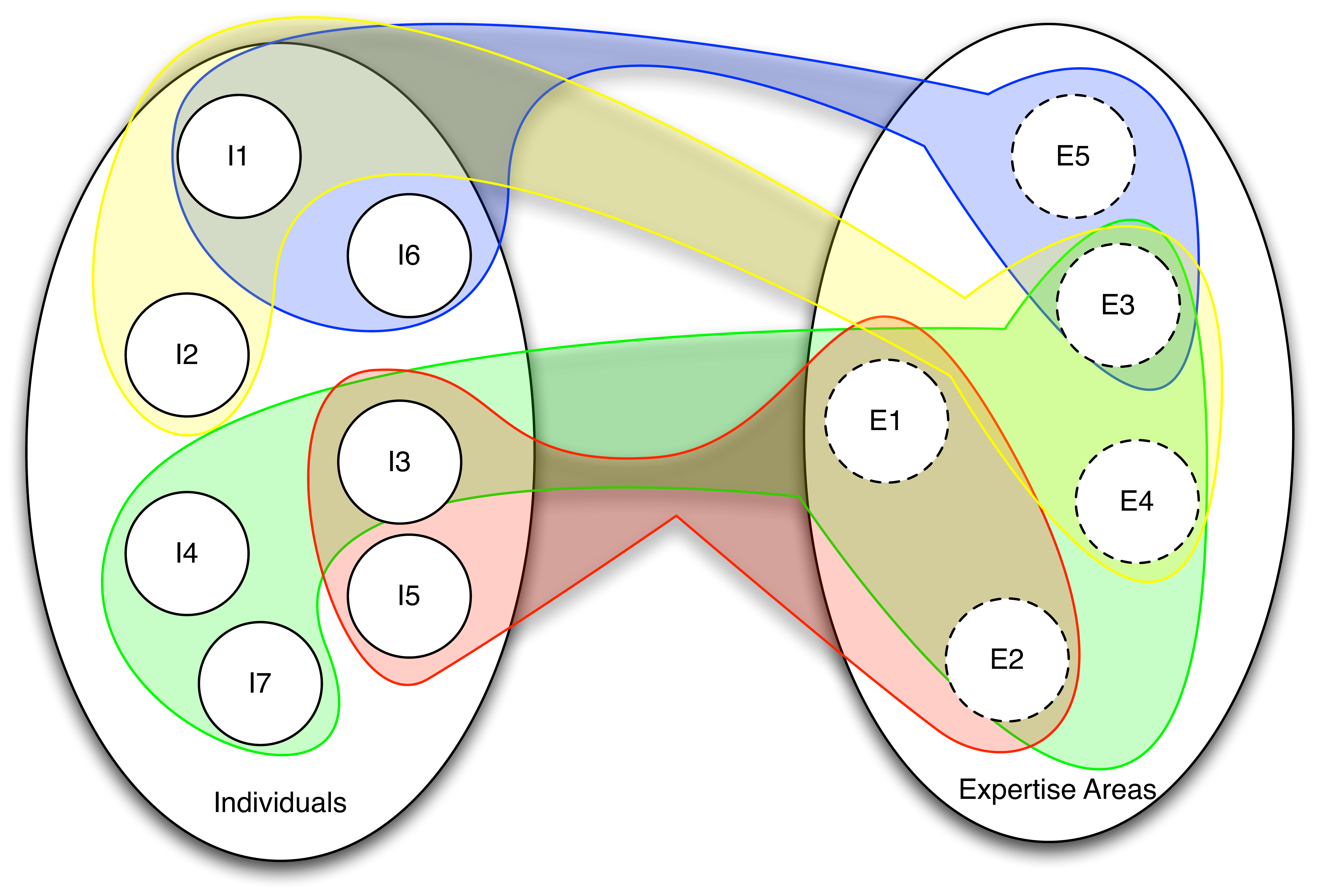}
\caption{The \emph{history graph} represents the past
    collaborations between individuals. Each collaboration (or team) is
    represented by a set of individuals, which is connected to a set
    of expertise areas. Such selected expertise areas are the competencies for which the individuals had been selected as team members.}
\label{fig:history}
\end{figure}

In order to model and algorithmically analyze the available information, it is necessary to capture and codify it using some well defined data structures and mathematical objects. Specifically, the SWAT team recommendation framework is based on a model characterized by three sets: individuals $I$, expertise
areas $EA$ and social dimensions $SD$. The elements of such sets are
captured using three graphs, namely the \emph{competence graph}, the
\emph{social graph} and the \emph{history graph} as shown in 
Figure~\ref{fig:competence}, Figure~\ref{fig:social} and Figure~\ref{fig:history}



The competence graph (Figure~\ref{fig:competence}) associates each individual with the expertise
areas in which s/he is competent and is defined as a labeled bipartite graph
of the form $(I, EA, E)$ where, as previously mentioned, $I$ is the
set of individuals, $EA$ is the set of expertise areas and $E \subseteq
I \times EA$ is the set of edges of the graph.
The label associated with each edge $(i, ea)$ is a real number $c \in
(0, 1)$ which is used to specify the degree of competence of the
individual $i$ in the expertise area $ea$.

Consider, for example, an individual named Alice. She is very good
at playing the piano but not so much at playing the guitar. Hence,
Alice participates in the competence graph in two edges. The first
edge is of the form $(Alice, piano)$ and we assign to it the label
$0.9$, showing her high competence in the area. On the other hand,
the label associated with the other edge -- $(Alice, sing)$ -- is
$0.3$, reflecting the fact that she is not a skilled singer. How
such information is derived depends on the nature of the underlying
data (to be elaborated later for a specific case study), and is
orthogonal to the abstract model in itself.

The social
  graph represents the (multiple) relationships existing among
  individuals. Formally, the social graph is a directed multigraph of
  the form $(I, SE)$, where $I$ is the set of individuals and $SE
  \subseteq I \times I$ is the set of edges connecting the
  individuals. 
  Each edge $se \in SE$ is labeled
  with 
  a tuple of the form $(d, s)$ where $d \in SD$ is the social
  dimension represented by the edge and $s \in (0, 1)$ is the strength
  of the social relationship that it represents.
  The social graph is a multigraph because two individuals may be
  connected by more than one social dimension, thus two individuals
  may be connected by more than one edge. Note that, given two
  individuals $i, j \in I$ the directed edges connecting $i$ and $j$
  must be labeled with different social dimensions. 
  Moreover, the social graph is directed because the strength of the
  social relationships between two individuals may differ according to
  the side from which we look at it. Obviously, it may be that the
  relationships defined by specific social dimensions are, as a matter
  of fact, symmetric. In such cases the relationship existing between
  two individuals is represented by means of two edges in opposite
  directions but associated with identical labels. 


As an example, let us consider a social network with only one
dimension. Hence, let us assume that $SD = \{\text{colleague}\}$.
Now, suppose that Alice and Bob work in the same company. Because of
that
they are connected by means of two edges. 
Say, there is a mechanism to define the strength of the
\emph{colleague} dimension based on a pair's distance in the
organizational hierarchy. Because of that, the labels associated to
both the previously defined edges is of the form $(\text{colleague},
x)$. If we expand the social graph adding a new dimension,
$\text{friend}$ for example, then we may add two new edges
connecting the nodes Alice and Bob. This time we assume that the
strength of the relationship is defined by the ``trust'' that each
individual has on the counterpart. Therefore, one may assume that
the label associated to the new edge $(Alice, Bob)$, which is in
addition to the one labeled with the \emph{colleague} dimension, is
$0.7$, reflecting the fact that Bob is seen as a good friend by
Alice, while the one associated to another edge between Bob and
Alice is $0.3$ reflecting the fact that Bob has a different opinion
of Alice.

Finally, the history graph represents past collaborations
(teams). Formally, it is an undirected \emph{bipartite
  hypergraph}\footnote{Definitions from Wikipedia: A bipartite graph
  (or bigraph) is a graph whose vertices can be divided into two
  disjoint sets $U$ and $V$ such that every edge connects a vertex in
  $U$ to one in $V$; that is, $U$ and $V$ are each independent sets. A
  hypergraph is a generalization of a graph in which an edge can
  connect any number of vertices.} of the form $(I, EA, T)$ where $I$
is the set of individuals, $EA$ is the set of expertise areas and $T
\subseteq I^{*} \times EA^{*}$, which means that each edge $t \in T$
is a tuple of the form $(I_{t}, EA_{t})$ where $I_{t} \subseteq I$ and
$EA_{t} \subseteq EA$.
Informally, the edges of the graph identify which individuals (the set $I_t$) collaborated on which expertise areas (the set $EA_t$).
Hence, for each relationship $t \in T$, $t =
(is, eas)$, we require that $|is| \geq 2$ and that $|eas| \geq |is|$.
This constraint is directly derived from our definition of team, that
defines as team a group of two or more individuals collaborating to
achieve a specific task. In turn, a task is defined as an objective
requiring a certain number of skills (more than one) to be
completed. Hence, each member of the team is selected to collaborate
contribute to at least one of the needed expertise areas in which s/he
is competent. Therefore, the expertise areas of any given edge are at
most as numerous as the individuals.
Other than keeping track of past
interactions of the individuals, the history graph helps the model to
track the evolution of the expertise area(s) in which the individuals
are involved. Such requirements originate from studies (such as
\cite{roth10}) where it is argued that more complex structures going
beyond egocentric representations (like the concept graph) are needed
to keep track of the presence and evolution of the set of users and
the set of concepts.

\subsection{Instantiating the model}
\label{sec:building_data}
\begin{figure}[htbp]
   \centering
   \includegraphics[width=.95\columnwidth]{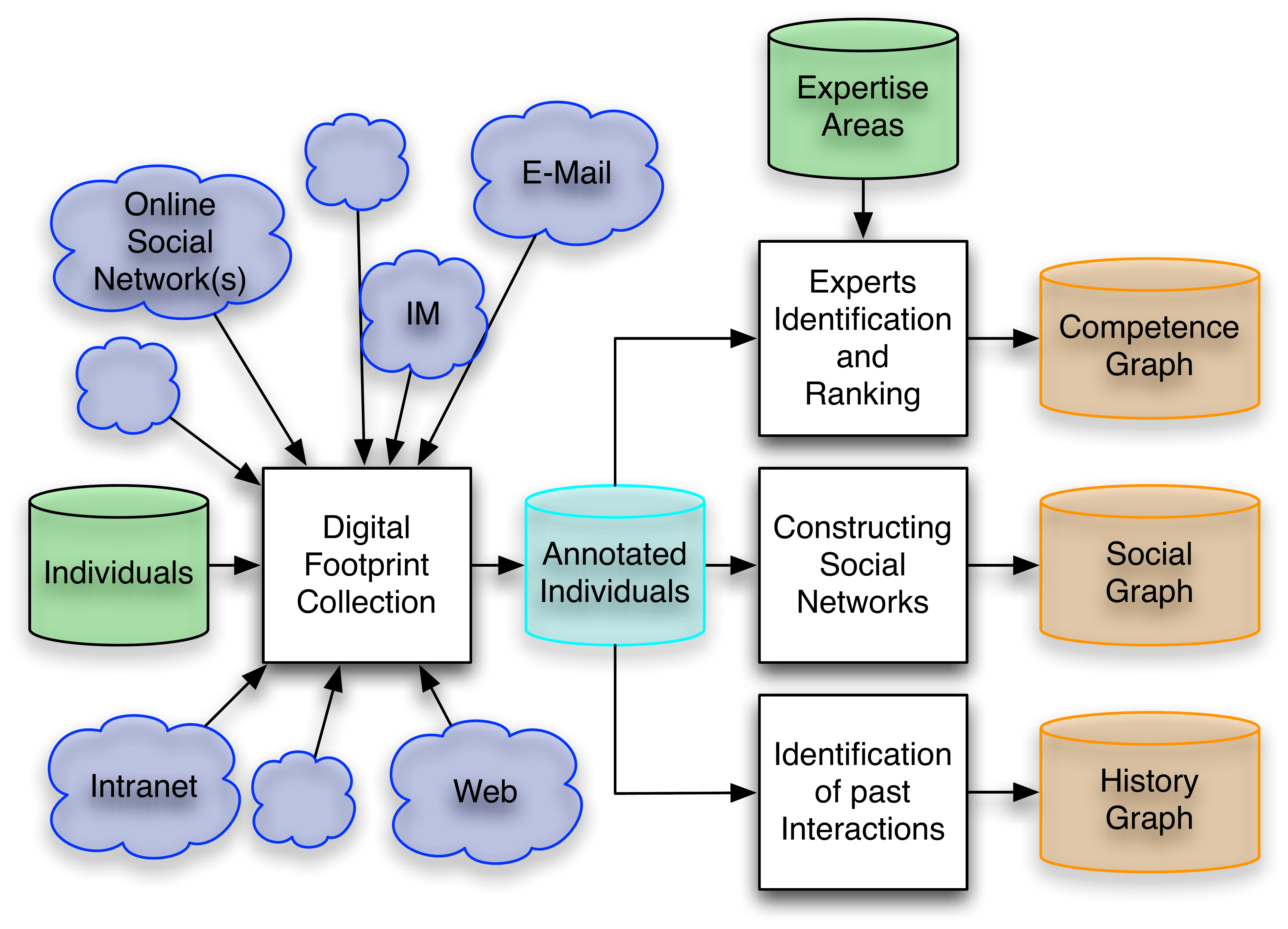}
   \caption{Data retrieval and generation of graphs.}
   \label{fig:highlevel}
 \end{figure}

In order to translate the model into something practicable, it needs to be instantiated with data relevant for a particular domain. The specifics of such an instantiation process will vary depending on the nature of the raw data and the domain. Nevertheless, at a high level, the instantiation process involves some basic modules that are illustrated in Figure~\ref{fig:highlevel}.

To start with, the set of individuals need to be identified, as well as the set of possible expertise areas and kinds of relations need to be enumerated in accordance to the specific domain. Furthermore, the relations (subsumption/similarity/etc.) among different expertise need to be identified.


By crawling (multiple) data sources - the web, online social networks, Intranet, etc. and applying necessary information retrieval mechanisms (such as entity extraction, graph mining, expertise identification, etc.) on the digital footprints so obtained, the annotated individuals (a person identified by a name along with all the information that the system is able to collect about her/him) corpus is built. A multitude of mechanisms may be deployed for these steps of crawling and annotation where the mechanisms may depend on the domain knowledge as well as data schema, etc. This corpus of annotated individuals is used to derive the three above mentioned graphs. Note that from the expertise areas associated to a past collaboration it may be possible to identify missing edges in the competence graph or incorrect association of expertise areas to the past collaborations, thus facilitating partial cross validation.



\section{A case study with academic research teams}
\label{sec:casestudy}
To demonstrate the applicability of the proposed framework, as well as determine the specificities of the high level modules, we have created a SWAT instance for academic teams. In Section~\ref{sec:implementation} we will present the implementation of the application along with some of the features that it provides. We first describe the process of harvesting and cleaning the necessary data, and some summary of the obtained corpus, which drives the application.

\subsection{Data harvesting}
\label{sec:harvesting}


Over time, we have created and demonstrated two versions of academic team recommendation systems, namely \cite{trecs} and \cite{swat_scbda}. The former \cite{trecs} was confined to NTU research staff, consisting of 1223 individuals, for whom the necessary records - complete list of publications and corresponding content (text), department, participation in funded projects, self-declared expertise areas, etc. were readily available (and was relatively clean) in a structured format from the university's research support office. In many corporate environments, employee profile information can similarly be used to populate the corpus.

Even with the ready availability of the necessary `digital footprints', there were several challenges - most prominent among these being the automatic detection and categorization of expertise (despite the self-declared information). A subproblem for categorization of expertise involves the establishment of relationships between expertise areas. Wikipedia categories was analysed to establish subsumption, similarity and synonymous relations \cite{trecs}.

In contrast to the very small, well-structured and readily portable datasets used in \cite{trecs}, our follow-up work \cite{swat_scbda} delves into capturing the data from the `wild' - namely from open web repositories such as DBLP, Academia.edu, etc. More precisely, the individuals are retrieved from the
DBLP database. From this dataset we were able to extract not only individuals -- used in the competence and social graphs -- but also to identify publications and venues -- used for the team graph.

This data in itself is however incomplete since it does not contain much
information about each individual and the relationships between
individuals (or publications), and information regarding expertise areas is very
ambiguous. Therefore, as mentioned in Section~\ref{sec:building_data}, we
had to crawl multiple online repositories in order to boost the information
contained in our dataset.

To expand the information about individuals, we took advantage of the
technophilia among academics, accordingly we implemented a set of wrappers
to retrieve the required data from public (academic) social networks
such as Academia.edu and Facebook. The latter
extracts the publicly available data and the private data if and only
if the legit data owner authorized us (see
Section~\ref{sec:implementation} for Facebook integration of SWAT). Using the information
contained in the named social networks we were able to populate the
social graph apart from the competence graph.

We also created crawlers
to extract information from public services that are strictly
related. Namely, we took advantage of academic indexing services
like Microsoft Academics and Google Scholar. As before, the extracted
data helped us in the identification of research interests and the
affiliations of the individuals. We also used the
Google Maps API in order to identify the geographic locations (country, region and
city) of the retrieved organization, in order to create a more accurate profile
of each individual. Such information can be used to create geographically confined or diverse teams, for instance.

Moreover, we expanded the information related to publications and
venues. The information contained in DBLP allowed us to identify the
services storing abstracts of several of the published
articles. Consequently, we created wrappers to extract public
information such as abstracts, topics and citation counts from
the following repositories --- IEEE Xplore, ACM Digital Library and Springer's Digital
Library. We also used the abstracts provided by CiteSeerX to complement
the dataset, filling the empty spots of abstract that our wrappers were
not able to retrieve from other repositories, and also to verify the
accuracy of the retrieved text.

Once we collected new information on the publications, we applied
topic extraction techniques to independently identify the topics of
each paper and we confronted the retrieved topics with the one
identified by the services that we crawled. Note that each of these steps (modules) can be realized by multiple possible techniques, and is agnostic with respect to the
topic extraction (a.k.a keyword extraction, keyphrase extraction,
concept extraction) technique used. In our case, we used a recent `home grown' Wikipedia data driven disambiguation technique \cite{li11} for the topic extraction process from the publications' titles and abstracts.


\subsection{Cleaning the data}
\label{sec:cleaning}

The dataset built from the various sources is often inaccurate and incomplete. Errors
are caused by wrong information stored in the crawled repositories, lack of a homogeneous schema, or well defined schema mappings across different sources, as well as due to
mistakes introduced by the wrappers themselves. Moreover, the topic
extraction techniques, while indeed quite accurate, may nevertheless lead to incorrect classifications sometimes.  Finally, even if the relationships between expertise areas and individuals that our tools
retrieve and derive are accurate at the time of retrieval, it may not remain the same over time (see
for example \cite{evolution} for an analysis of the evolution of authors profile).

Thus, the algorithmic mechanisms of deduction in our system are complemented with crowdsourcing tools, which take inputs from the end-users in order to rectify and learn new information. The integration of such tools in SWAT is described in Section~\ref{sec:implementation}.

\subsection{Statistics of the dataset}
\label{sec:statistics}

We briefly present some statistics regarding the harvested data. Table~\ref{tbl:statistics} presents the
current size of the collected data. Note that such statistics are most
likely outdated as the wrappers are constantly retrieving new data,
both extracting it from new sources -- by means of newly deployed
wrappers -- or updating the existing data from new database dumps
being released in the existing sources.

\begin{table}[htb]
\centering

\begin{tabular}{| c | c |}
\hline
\textbf{Statistic} & \textbf{Value}\\
\hline
\hline
Individuals & 996,717\\
\hline
Concepts  & 552,962\\\hline
Teams & 2,940,631 \\\hline
Average connections/individual & 7.2732 \\\hline
 Average individuals/team & 3.1180\\\hline
 Max individuals/team & 119\\\hline
Organizations & 2,524\\\hline
Countries & 80 \\\hline
\end{tabular}
\caption{Statistics of SWAT's knowledge base.}
\label{tbl:statistics}
\end{table}

\begin{figure}[htb]
\centering
\includegraphics[width=.95\columnwidth]{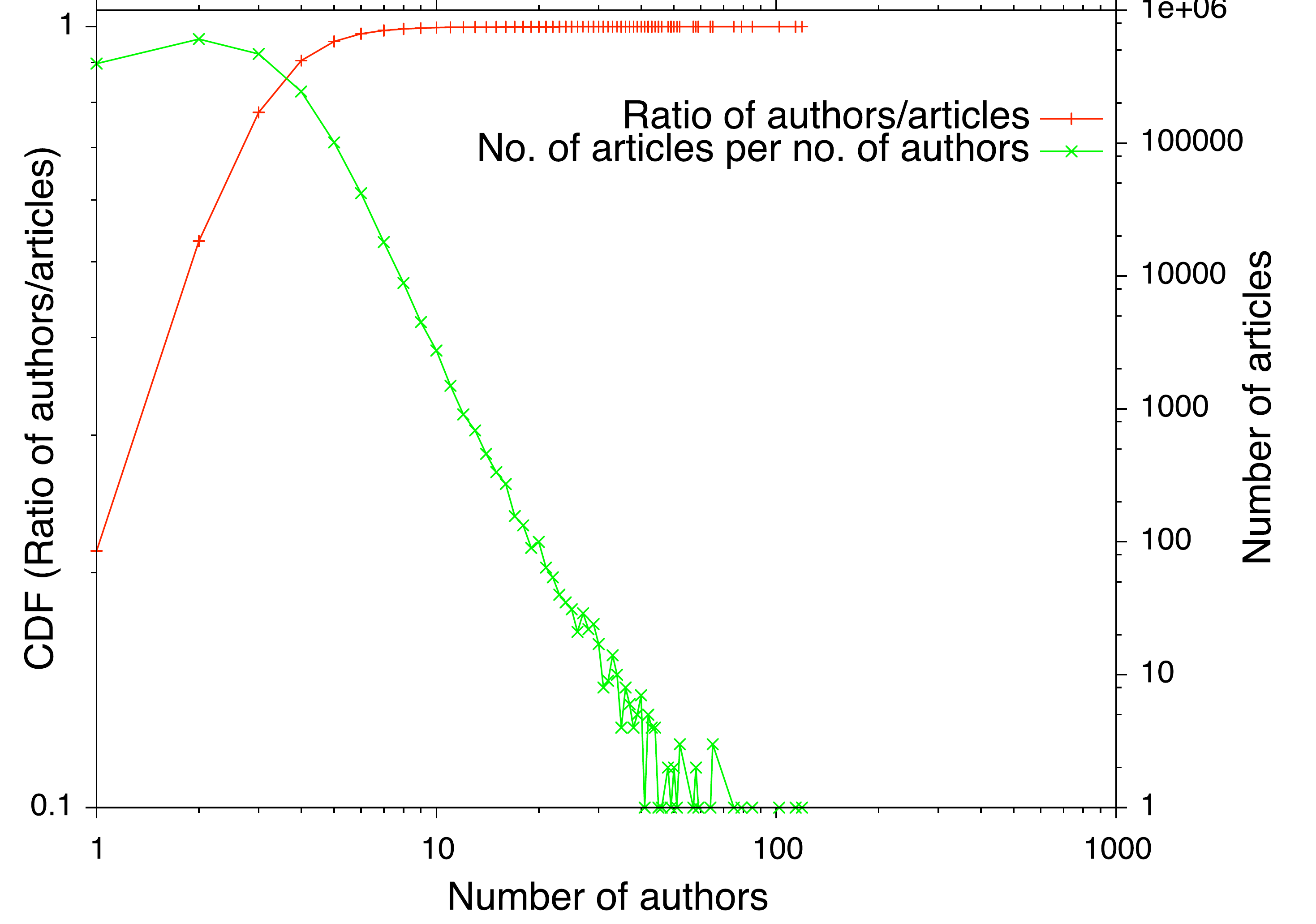}
\caption{Cumulative distribution of the number of authors per article
  and number of articles per number of authors.}
\label{fig:nauthors}
\end{figure}

From the collected data, we can revalidate the claims of other
studies \cite{grossman02,newman04}, particularly that there is a
growing trend of team work and larger teams.
Figure~\ref{fig:nauthors} shows that 21\% of the papers have only
one author, while 99\% of the papers have at most 8 authors. It is
more common to publish articles with two, three or four authors.
Moreover, one may notice that the number of articles with one author
is less than the number of articles with five-seven authors.
Finally, we notice that the articles with more than ten authors are
quite rare, nevertheless we identified articles with even more than
hundred authors.

\begin{figure}[htb]
\centering
\includegraphics[width=.95\columnwidth]{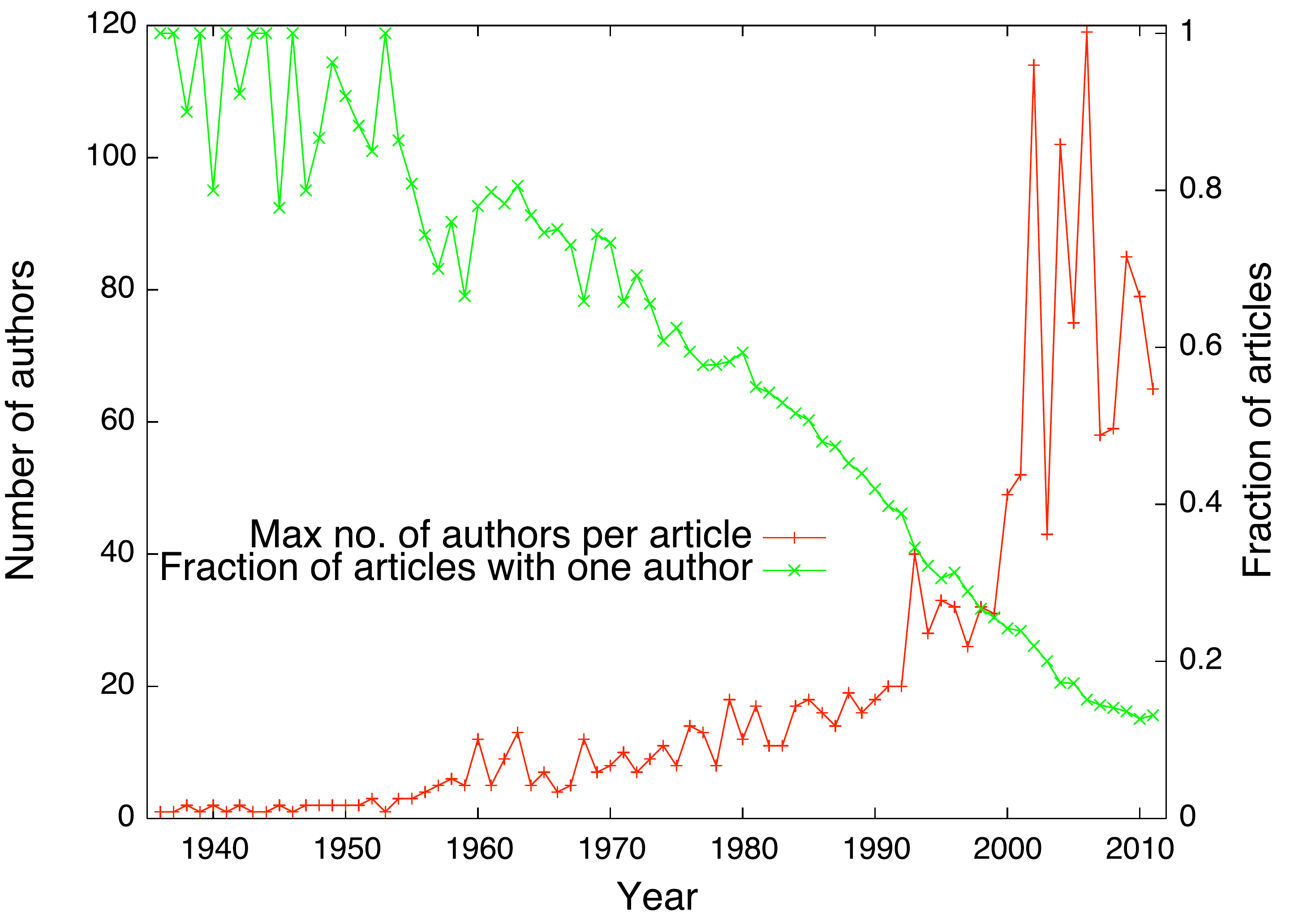}
\caption{Evolution of the number of authors per article.}
\label{fig:nauthors_years}
\end{figure}

The data presented in Figure~\ref{fig:nauthors} refer to the entire
dataset archived in DBLP from 1936 to 2011 (we removed 2012's
articles from this evaluation because the articles published are not
yet fully recorded in DBLP). Therefore, we also analysed the
evolution of the distribution of authors per article over this time
interval. As expected, Figure~\ref{fig:nauthors_years} shows that
the percentage of articles with only one author is decreasing over
time. Similarly, the maximum number of authors per article is slowly increasing,
reconfirming perceptions about a growing trend of collaborative and
team works in academia.

\section{Team formation and ranking}
\label{sec:team_formation}

Acquiring, merging and cleaning the necessary data and storing it in a
suitable fashion is in itself nontrivial, and involves many steps
summarized above. Once the suitable data is in place, leveraging it
for understanding and recommending teams involves another complex set
of tasks.

For a given objective, the necessary set of expertise needs to be
identified, followed by the identification of a group of individuals
who together cater to the multiple skills requirement, where a single
individual may be competent in several necessary skills. Such a group
catering all the skill needs can be called a team. However, many
possible combinations would typically exist. Hence, it is desirable to
quantify the possible teams' suitability, possibly subject to certain
characteristics.

In the current SWAT realization, we support four metrics - but other
measures can readily be integrated to extend it. The currently used
metrics have been carefully tested \cite{swat_scbda,trecs} and are
primarily inspired by several studies that analyzed team cohesion and
dynamics to determine what are important parameters for a team's
formation and success
\cite{oss,collabonline,gameteam,wikiteam,socinfo11}.

The first two metrics, initially described in \cite{trecs}, are 
\textbf{competence score} and \textbf{social cohesiveness score}. The
competence score is a measure of the competence of the team users with
respect to the skills (expertise areas) required to achieve an 
objective. It is not a simple algebraic sum of the members' competence values
-- the label of the edges of the competence graph connecting the
members (individuals) to the expertise areas -- and the metric can be
tuned to emphasize different aspects. As an example, it is possible to
tweak the metric so that it computes an average of the different
competence values to privilege teams where the members have, more or
less, equal competence. On the other hand, it is possible to let the
metric to identity only the higher competence value, to privilege
those teams with at least a member highly competent in some skills.

The social cohesiveness score is a metric that measures the ``social''
characteristics of a team. This metric is relevant if one assumes that
the team members work better together if they share some social
relationships. In SWAT, we use a modified \emph{clustering
  coefficient} measure to determine the proximity among team
members. Note that the underlying social graph we use in SWAT is
multidimensional, and thus paths among individuals are determined by
exploring several of these relations, and it may happen that two
individuals are not connected through a work-related path -- for
example by means of a series of coauthors -- but are connected if
another kind of relationship is taken into account, say friendship.

The latter two metrics, identified and used in \cite{socinfo11} and
\cite{swat_scbda} respectively are \textbf{team user repetition} and
\textbf{team concept repetition}. Both the metrics are used based on
the observation that if members of a team worked together in the past
then there are better chances that the given team would work
successfully again.


  The team user repetition metric counts the number of past teams --
  represented in the model by means of the team graph -- whose members
  are a subset of the members of a given team.
  Thus, it measures the likelihood of the individuals to work together
  effectively. On the other hand, the team concept repetition metric
  measures the similarity, in terms of concepts instead of members,
  between the current team and past teams. By doing so, the metric
  measures the likelihood of the current team to work effectively on
  the concepts required by the current objective.


\section{Implementation and evaluation}
\label{sec:experiment}

While there is arguably no objective truth, and many sources of error - incomplete and erroneous data, imprecise building blocks (say, for expertise identification, graph mining, etc.), and no means to quantify even the optimality of the manner in which these building blocks are composed together, a team recommendation system can still be qualitatively benchmarked by looking at the kind of results it churns out, and even if the results are sub-optimal (whatever that may mean in the absence of any ground truth), the system can still be useful as a decision support tool. We have accordingly implemented the framework as a web application, which can be used to shortlist academic experts for participation in grant proposals and projects, or to find reviewers and TPC members, etc. The prototype is available at \texttt{http://sands.sce.ntu.edu.sg/SWAT/}.

\subsection{Implementation}
\label{sec:implementation}

The web application has been implemented using Java 5 for the
backend and GWT (Google Web Toolkit) 2.4 for the user interface. It
stores the data in MySQL databases and uses the Solr search platform
\cite{solr} to speed up queries. It is also fully integrated with
Facebook, allowing users to access more advanced services if
registered. The application works also in stand-alone mode,
providing ``only'' the team recommendation service to casual users.
In fact, when a new user access the web site s/he may choose to
proceed as a guest or as an authenticated user.

If the user accepts to register within SWAT, the system will guide the
user through a series of steps to her/his accurate identification
among the individuals of the dataset. If the user is not an
existing individual captured automatically from the crawled data-set, s/he will be provided a wizard interface to create a
new entry in the dataset. Once the user is registered, the backend
will activate the Facebook's wrapper to retrieve part of the user's
profile to extend the already collected information. In particular,
the wrapper will retrieve friendship information to populate another
dimension of the social graph.

The identification of the best experts for a specified expertise
area is handled by the Expert Selection module. Such module can be
accessed, as previously mentioned, also by unregistered users.
Similar (possible) expertise areas will be suggested while the user
is typing the desired query. For instance, if a user types
``social'' then the system will suggest ``Communication Technologies
And Social Change,
 Online Social Network,
 Social Cognition,
 Social Network Analysis,
 Social Psychology, etc.''.

\begin{figure*}[htbp]
  \centering \subfloat[The user registration interface allows a new user to associate her/his Facebook account to one of the individuals identified by our system.]{
\includegraphics
[width=.94\columnwidth]
{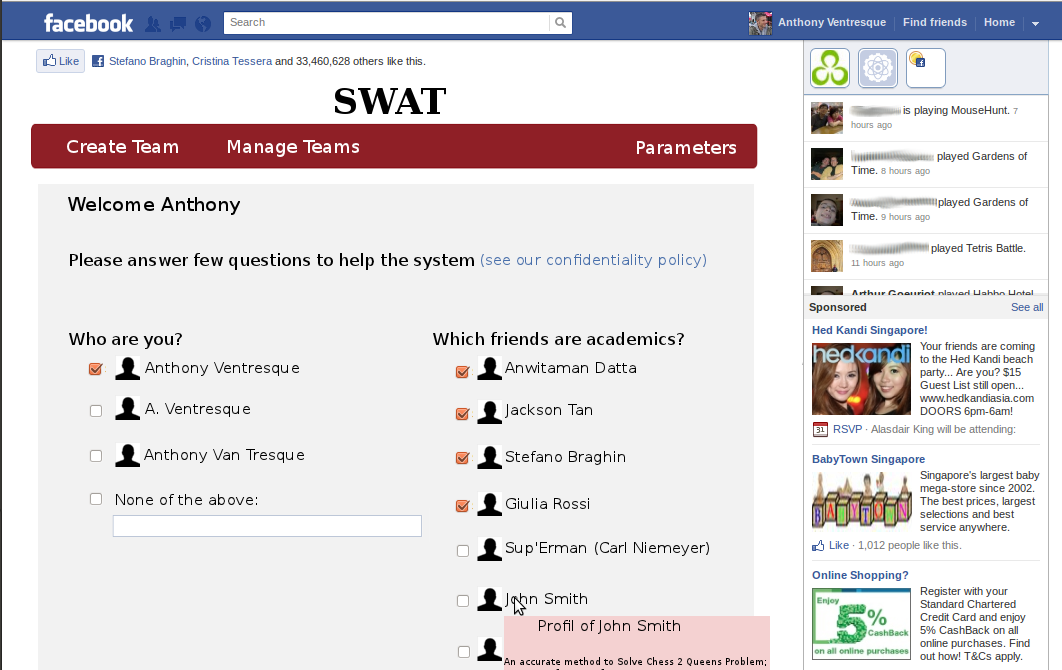}
}\qquad \subfloat[The team selection module
    allows the user to evaluate the recommended teams. Through this
    interface the user is able to visualize the team scores according
    to different representation (A), re-compute the recommended teams upon having adjusted the weight of the different metrics (B) or to contact a recommended team (C).]{
\includegraphics
[width=.94\columnwidth]
{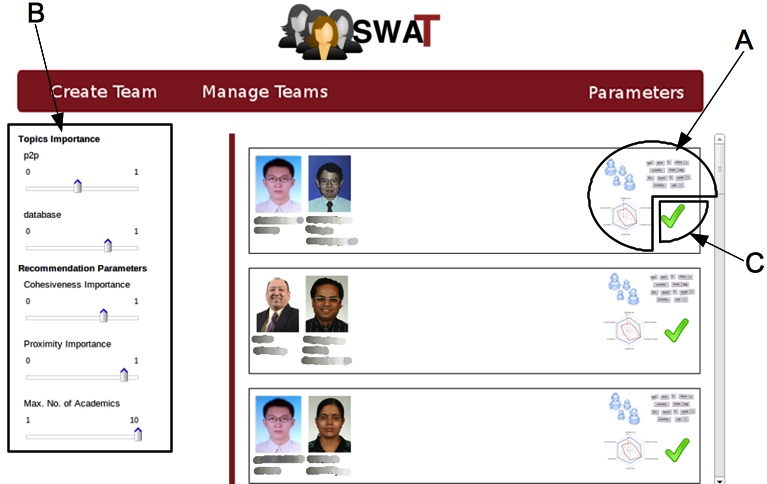}
}\\
  \subfloat[The graphical representation of the relationships between
  team members. The label of the edges connecting author nodes
  represents the length of the shortest path on the social graph; the
  label on the edges connecting an author node to a concept represents the competence value of that specific user for that specific concept.]{
\includegraphics
[width=.94\columnwidth]
{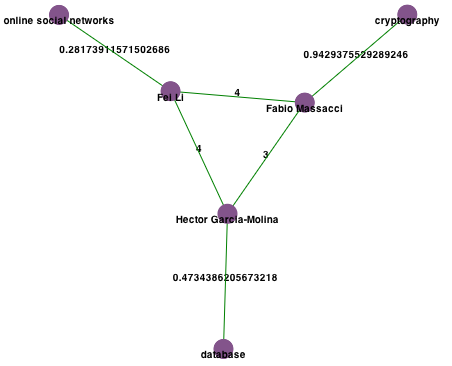}
}\quad \subfloat[The graphical representation of the team's scores.
The axes of the \emph{radar chart} represent the metrics described
in Section~\ref{sec:team_formation} and other information about the
selected team.]{
\includegraphics
[width=.8071\columnwidth]
{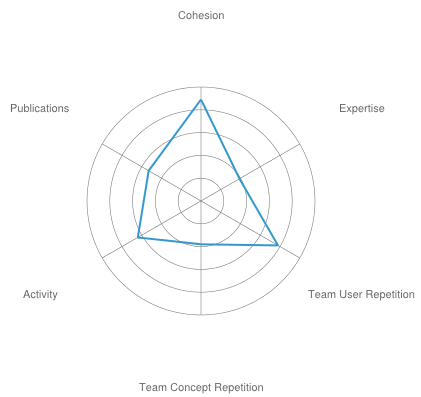}
} \caption{Some screenshots of the SWAT web application.}
\label{fig:biggraphics}
\end{figure*}

Such suggestions are created from similarity of Wikipedia
categories. Once the user has confirmed the input, the system will
retrieved a list of users who have expertise relevant or similar to
the input. These are all based on the profile, publications, keywords
and relevant information retrieved from various sources described in
Section~\ref{sec:harvesting}. Once such a list of experts is retrieved, the
user will be able to review the profile information of each expert and
to have a preview of her/his egocentric network.

After the desired concepts are all selected, the Team Creation module
computes all the possible teams and present the results to the user.
Teams will be formed based on the retrieved experts (which we remind
are individuals connected to expertise areas in the competence graph)
from each concept. The computed teams are ranked according to the
scores computed by the metrics presented in
Section~\ref{sec:team_formation}, and weightage of these metrics as preferred by the user. The user interface also allows the user
to navigate the characteristics of each recommended team through
different views such as member details, where the team members profile
are summarized along with their publication, affiliations, etc.
Moreover it is possible for the user to access both a graphical
and textual description of the scores computed on the team, to have a
better feeling of the characteristics that let the system identify
such team as a good one. The user is also able to visualize
a graphical representation of the social relationships existing
between the team members and the relationships existing between the
members and the required expertise areas. 
Finally, the user can manually compose a team or edit the team
members, and observe the various metrics corresponding to the team
composition.

The system provides a Contact Team interface for users to contact the
selected team members. The interface will prompt whether the team members can be contacted via
the Facebook messages (this happens when the user to be contacted is registered with SWAT) or the user has to do it manually via other means (say, email).

If the user is registered then s/he will be able to keep track of
the contacted teams through the Team Management module. Such a module
also provides an interface for users to accept, reject or
conditionally accept invitation to be members of other
teams/projects. Furthermore, the module works as an additional communication
channel between the different team members through a messaging system.

The web interface provides also the data cleaning tools mentioned in
Section~\ref{sec:cleaning}. Namely, a registered user will be allowed
to correct and update the collected information, particularly her/his own profile including the list of publications and the expertise areas in which s/he is
interested. Moreover, both registered and unregistered users can access crowdsourcing tools such as tests to verify the accuracy of the
topic extraction techniques, the correctness of the collected data, among others. The information collected through such tools do not directly
modify the knowledge base but are presented to the application's
administrators who have to approve them. This is akin to Google scholar's approach. As a future work we are going to introduce a reputation-based access control mechanism that will allow
registered users to operate as administrators and to modify more and
more sensitive data (somewhat similar to Wikipedia's approach of using trusted moderators).


\begin{figure}[htb]
  \centering
  \includegraphics[width=.95\columnwidth]{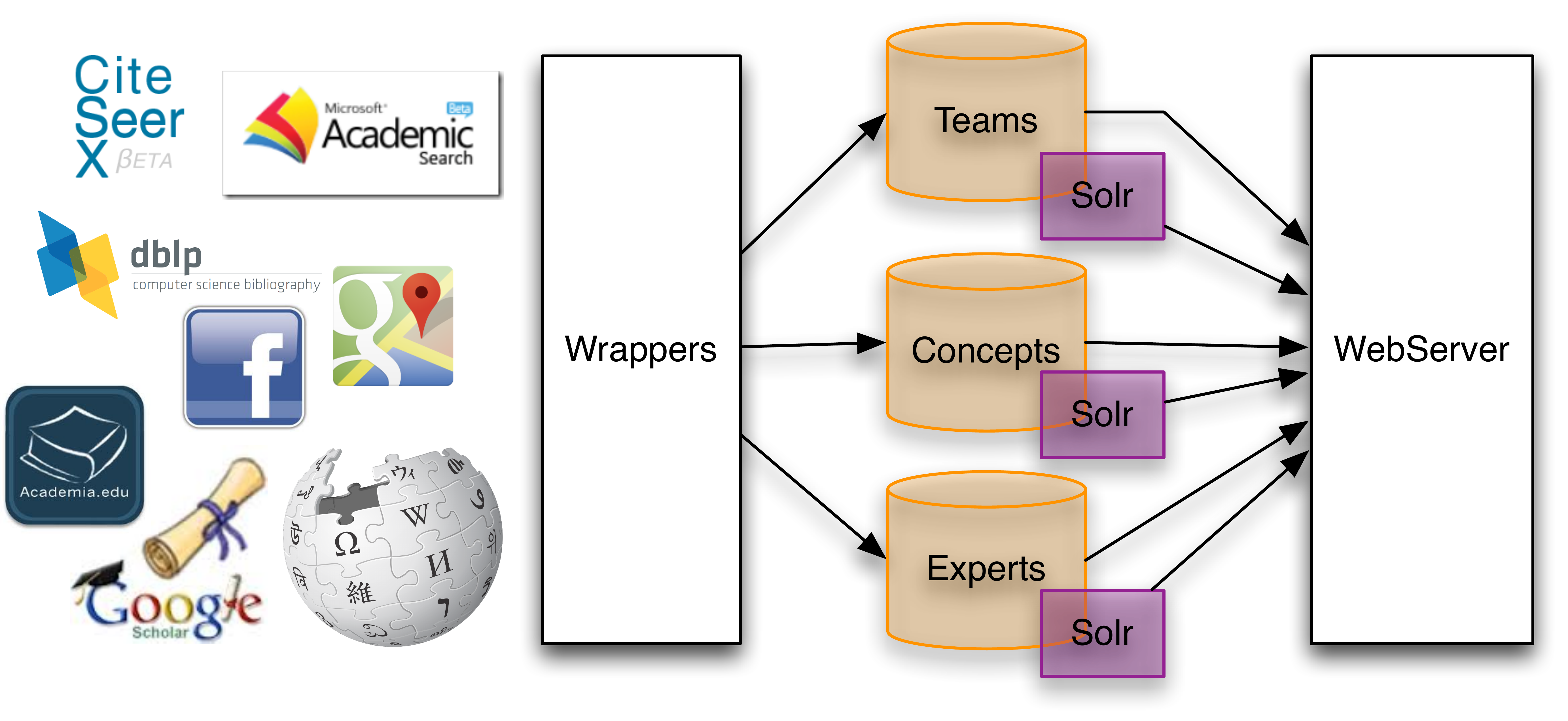}
  \caption{The architecture of the implemented web application}
  \label{fig:webapp}
\end{figure}

\subsection{Empirical evaluation}

We evaluated the performances of the implemented web
application. 
We run the experiments on the dataset described in
Section~\ref{sec:harvesting}.

\begin{figure}[htb]
\centering
\includegraphics[width=.95\columnwidth]{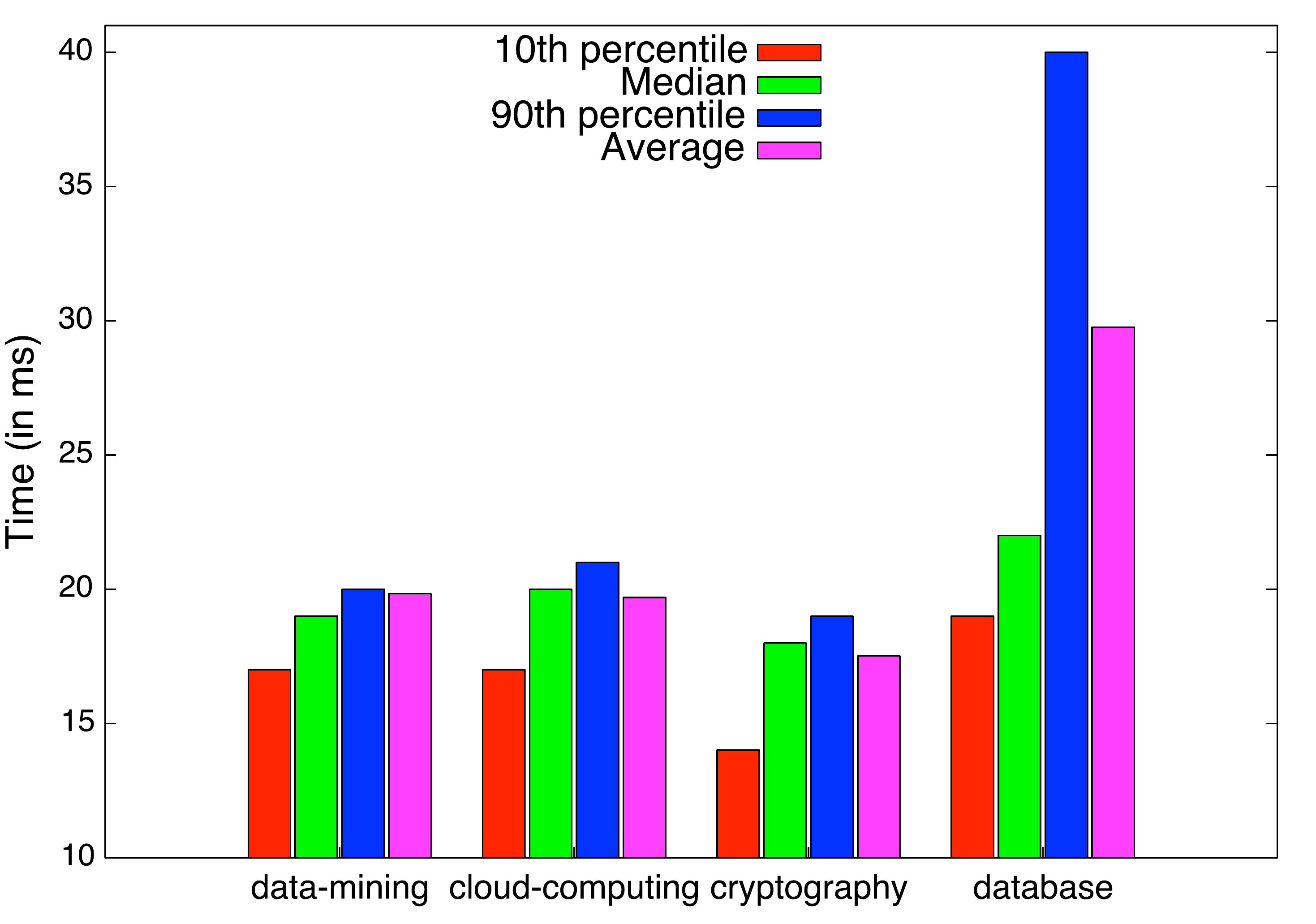}
\caption{Time required to extract from the databases the 20 top experts given an expertise area.}
\label{fig:expert-selection}
\end{figure}

First of all, we evaluated the time required to identify the top
twenty experts for some given expertise areas. Because of the original
source of individuals (DBLP) in our corpus, a significant portion of the records are
database-related papers. Hence, we evaluated the retrieval of experts in
related areas -- namely \texttt{data mining}, \texttt{cloud
  computing}, \texttt{cryptography} and \texttt{database} -- to show
the efficiency of the indexing services. As shown in
Figure~\ref{fig:expert-selection} the time required to extract the top
twenty experts is roughly 20 milliseconds, with the exception of the
expert area \texttt{database} which required longer, 30 milliseconds in
average. Hence, we are able to claim that even if the system
handles a large number of data (see Table~\ref{tbl:statistics}) it
is very responsive.

\begin{figure}[htb]
\centering
\includegraphics[width=.95\columnwidth]{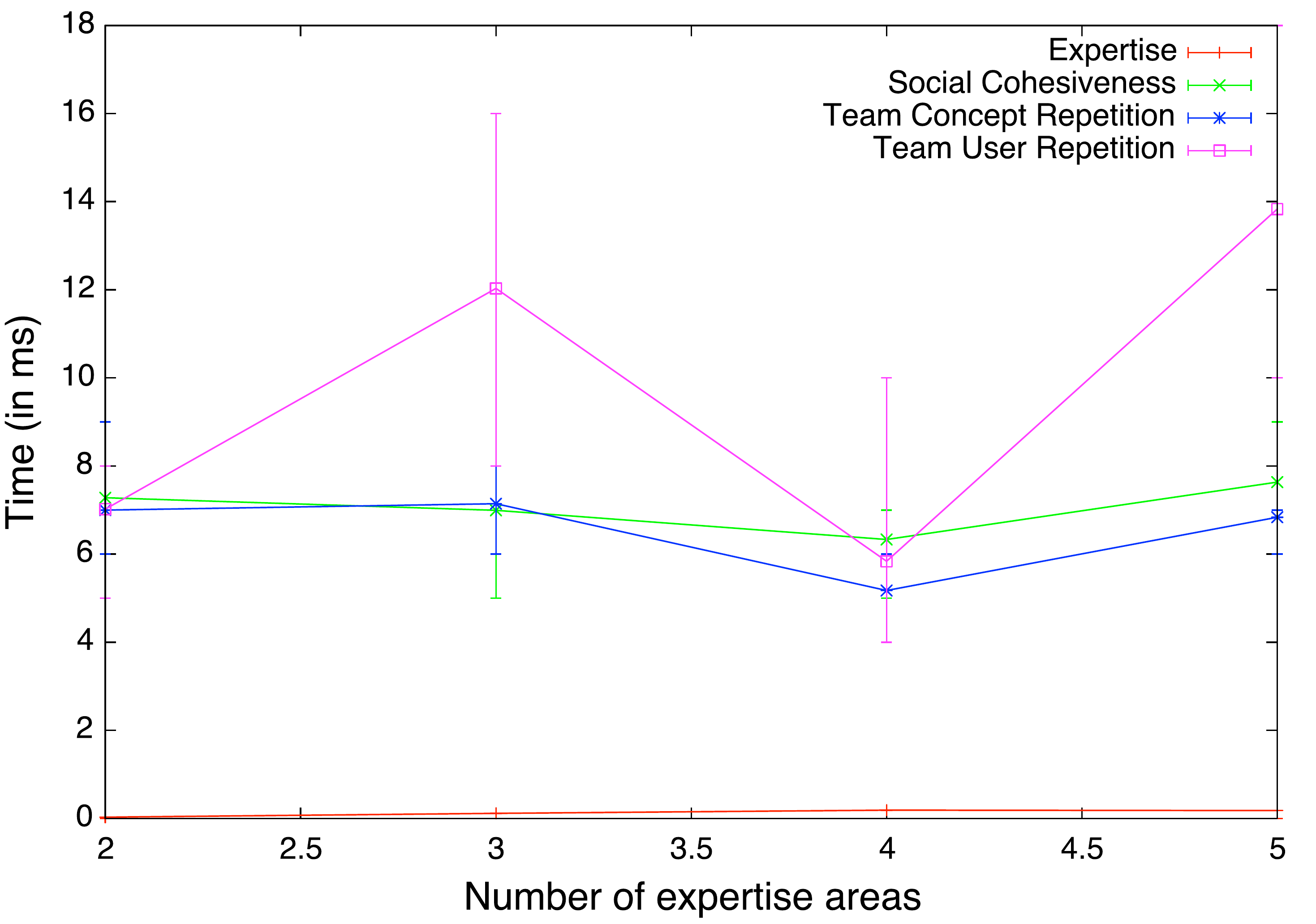}
\caption{Time required to compute each metric.}
\label{fig:team-metrics}
\end{figure}

After that, we evaluated the time required to compute team
recommendation. To do that, we measured the time required to present
to the user the twenty top teams for tasks requiring from two to five
expertise areas. First of all, we observed the time required to compute
the metrics of each team. Figure~\ref{fig:team-metrics} shows that the
time to compute SocialCohesivenes and Team ConceptRepetition metrics
is more or less constant -- and around 5 milliseconds. On the other
hand, the Expertise metric is determined practically immediately, taking less than
a millisecond to compute. The main reason for such a behavior is because
the Expertise metrics does not require to access the databases anymore to
retrieve any other information. All the data required for its computation
are already present in the User objects. In contrast, the
TeamUserRepetition accesses repetitively the database,
hence it is slowed down by concurrent threads and by the
overall load level of the database server.

\begin{figure}[htb]
\centering
\includegraphics[width=.95\columnwidth]{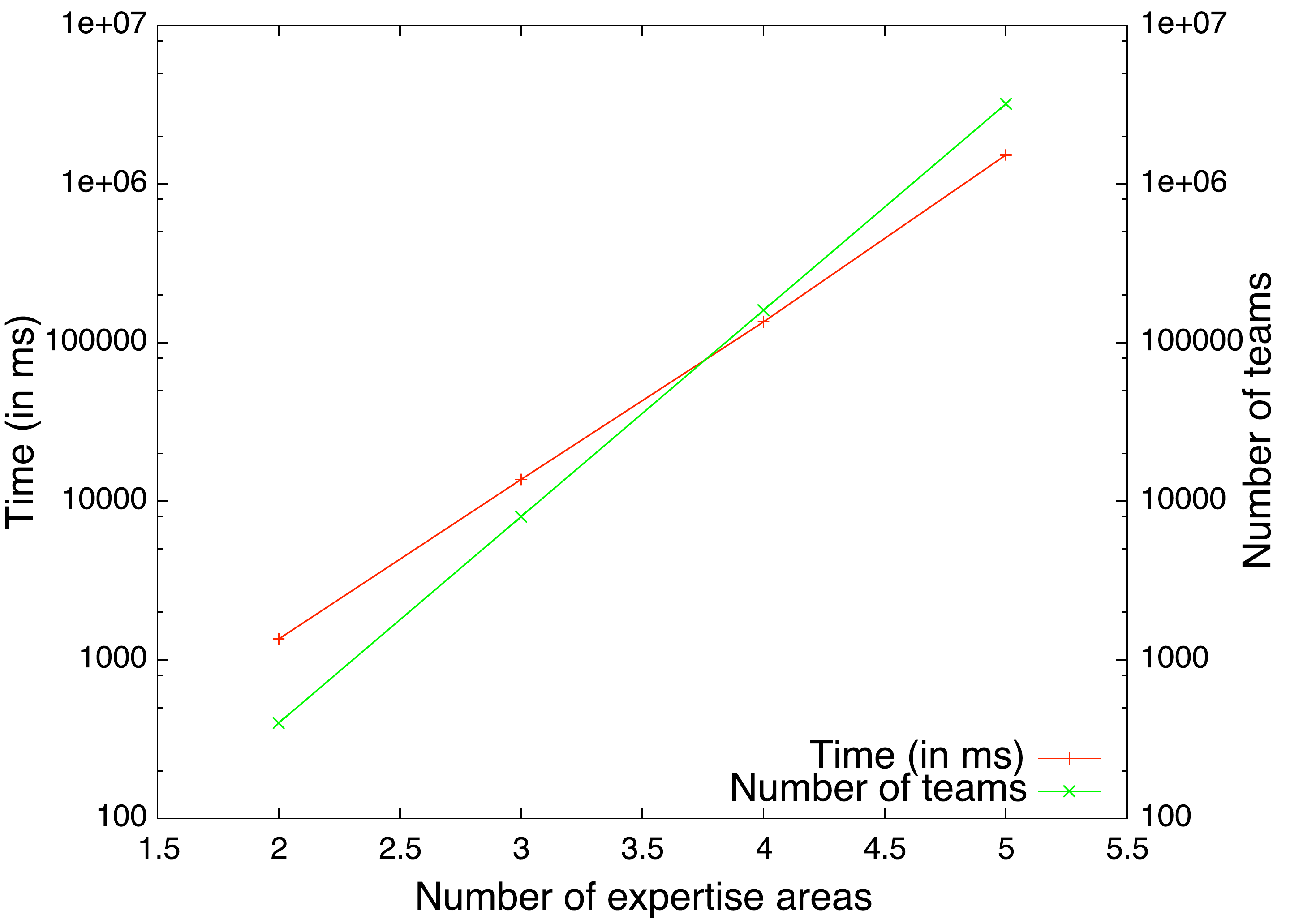}
\caption{Time required to generate the team suggestionm, y axis uses a logarithmic scales.}
\label{fig:team-all}
\end{figure}

Finally, Figure~\ref{fig:team-all} presents the execution time to
present to the user the top teams for the given expertise areas. As
one may notice, the time required depends on the number of expertise
area, as for each area 20 new users are involved in the process and,
therefore, the number of teams to be created and measured is increase
by a factor of 20. Thus, the time to compute the suggestions is linear
to the number of possible teams and, therefore, exponential to the
number of expertise areas.

\section{Concluding remarks}
\label{sec:conclusion}

In this paper we presented SWAT, a comprehensive framework for
multidisciplinary team recommendation. We informally presented the
model and the algorithms used to identify potential teams for the
completion of specific objectives. We also presented a specific
instance of SWAT for the suggestion of academic teams. The knowledge
base on which the presented instance works have been built by
retrieving the various information from different repositories
available on the web.

Each of the individual modules which have been composed together to
build SWAT can arguably be improved, and pose standalone research
challenges. Likewise the data that is mashed together may be boosted
and improved with further sources, as well as better data harvesting
and cleaning techniques. Taking into account the existing
commitments of individuals, and their consequent availability for a
project is also desirable. These comprise the obvious steps to
improve the SWAT application.

In addition to the improvements of the existing modules,
identification and incorporation of further new metrics to quantify
team characteristics, as well as integrating additional (new as well
as existing third party) tools for a more seamless collaboration,
and harnessing detailed digital footprints directly from these
collaboration tools themselves are some other aspects that should be
explored.


\bibliographystyle{abbrv}
\bibliography{swat}

\end{document}